\documentclass[prb,showpacs,twocolumn,floatfix]{revtex4}
\usepackage{graphicx}
\usepackage{dcolumn} 
\usepackage{bm}

\begin{document}
\def\rhov{{\mbox{\boldmath{$\rho$}}}}
\def\tauv{{\mbox{\boldmath{$\tau$}}}}
\def\Lambdav{{\mbox{\boldmath{$\Lambda$}}}}
\def\Thetav{{\mbox{\boldmath{$\Theta$}}}}
\def\Psiv{{\mbox{\boldmath{$\Psi$}}}}
\def\chiv{{\mbox{\boldmath{$\chi$}}}}
\def\Phiv{{\mbox{\boldmath{$\Phi$}}}}
\def\sigmav{{\mbox{\boldmath{$\sigma$}}}}
\def\xiv{{\mbox{\boldmath{$\xi$}}}}
\def\oh{{\scriptsize 1 \over \scriptsize 2}}
\def\ot{{\scriptsize 1 \over \scriptsize 3}}
\def\of{{\scriptsize 1 \over \scriptsize 4}}
\def\tf{{\scriptsize 3 \over \scriptsize 4}}
\title{A SYSTEM EXHIBITING TOROIDAL ORDER}

\author{A. B. Harris}

\affiliation{Department of Physics and Astronomy, University of
Pennsylvania, Philadelphia PA 19104}
\date{\today}
\begin{abstract}
A two dimensional system of discs upon which a triangle of
spins are mounted is shown to undergo a sequence of 
interesting phase transitions as the temperature is lowered.
We are mainly concerned with the `solid' phase in which bond
orientational order but not positional order is long ranged.
As the temperature is lowered in the `solid' phase, the first
phase transition involving the orientation or toroidal charge of
the discs is into a `gauge toroid' phase in which the product of
a magnetic toroidal parameter and an orientation variable (for the
discs) orders but due to a local gauge symmetry these variables
themselves do not individually order.  Finally, in the lowest
temperature phase the gauge symmetry is broken and toroidal
order and orientational order both develop. In the `gauge toroidal' 
phase time reversal invariance is broken and in the lowest temperature
phase inversion symmetry is also broken.  In none of these
phases is there long range order in any Fourier component of the
average spin. A definition of the toroidal magnetic moment $T_i$ of the
$i$th plaquette is proposed such that the magnetostatic interaction
between plaquettes $i$ and $j$ is proportional to $T_iT_j$. 
Symmetry considerations are used to construct the magnetoelectric free
energy and thereby to deduce which coefficients of the
linear magnetoelectric tensor are allowed to be nonzero. In none of
the phases does symmetry permit a spontaneous polarization.
\end{abstract}
\maketitle

\section{INTRODUCTION}

\begin{figure}[ht]
\begin{center}
\includegraphics[width=8.0 cm]{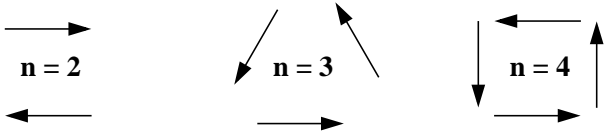}
\caption{\label{FIG1} Toroidal plaquettes with 2, 3, or 4 moments.}
\end{center}
\end{figure}

The theoretical analysis of toroidal ordering in 
electric[\onlinecite{PRPT,ETOR,PROS}] and 
magnetic[\onlinecite{PRPT,NAS1,NAS2}]
systems has recently been investigated.
Examples of such states in which plaquettes of spins assume a chiral
configuration have been known for some time.[\onlinecite{WEN}]
However, it is widely believed that
toroidal magnetic order should always be subservient to the primary
order parameter, a Fourier component of the average spin.
Here we address the possibility of defining a toroidal order parameter
for a system in which it is the {\it primary} magnetic order parameter.
The major problem is to identify a situation in which there is
ferrotoroidicity but there is no nonzero Fourier component of average
spin.  In this paper we consider a two dimensional system of toroidal spin
plaquettes, such as those shown in Fig. \ref{FIG1}, which exhibits
the desired behavior.

\section{THE MODEL}

\begin{figure}[ht]
\begin{center}
\includegraphics[width=8.0 cm]{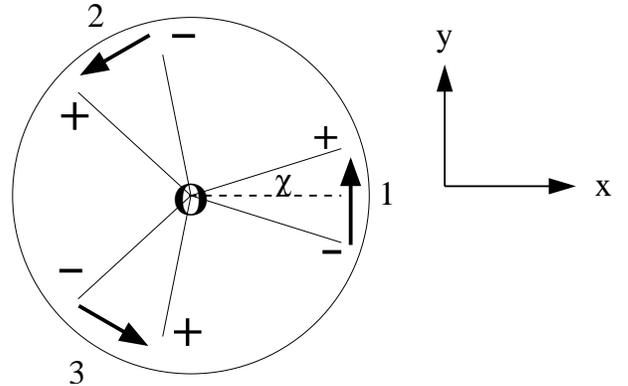}
\caption{\label{FIG2} A plaquette with three spins
in one of their dipolar ground states.  Each magnetic dipole is
represented by a pair of charges $\pm Q$ whose positions are
fixed by the angle $\chi$.}
\end{center}
\end{figure}

We consider a system of microscopic
circular discs (confined to lie in the $x$-$y$ plane)
which contain three spins in a triangular configuration as in
the center panel of Fig. \ref{FIG1}. 
Each spin has a large single ion anisotropy so that it is aligned 
either parallel or antiparallel to its local axis fixed in the plane
of the disc, as shown in Fig. \ref{FIG2}. The intraplaquette dipolar
interactions are strong enough so that at temperatures of interest
the spins in each plaquette come to thermal equilibrium in one of
the two degenerate ground states
as shown in Fig. \ref{FIG1}.  The magnetic dipole moment of
each spin is mimicked by a pair of opposite charges ($\pm Q$),
as shown in Fig. \ref{FIG2}.  If $r$ is the distance of the 
charges from the center of the plaquette, then the magnitude of
the dipole moment is $p=2Q r \sin \chi$.  The orientation of
the disc is defined by the angle $\phi$ between the $x$-direction
and the dashed line fixed on the plaquette.  

We now calculate the interaction energy $V_{AB}$ of plaquette A
whose center is at the origin and plaquette B whose center is at
${\bf R} \equiv (X,Y) \equiv (R \cos \Psi , R \sin \Psi)$.  To
calculate $V_{AB}$ we invoke the multipole expansion of the electrostatic
energy (which is proportional to the desired magnetic dipole-dipole
energy) between the charges on the two plaquettes.  This energy
is a sum over terms of the form 
$(1/R) F_n(\phi_A, \phi_B, \Psi) z^n$, where $z \equiv r/R$.
The positions are functions of $r\exp(\pm i \chi)$ and the 
symmetry of the plaquette indicates that the result can only 
involve factors of $r^3\exp(\pm 3i \chi)$.  Also this term must be
odd in both $\chi_A$ and $\chi_B$.  Accordingly, the lowest order
nonzero term is of order $z^6/R$.  So at this order
\begin{eqnarray}
V_{AB} &=&
\Lambda(\phi_A, \phi_B, \Psi) Q_A \sin(3\chi_A)Q_B \sin(3\chi_B)r^6/R^7 \ ,
\nonumber \\ &&
\end{eqnarray}
where $Q_A$ ($Q_B$) is the amplitude of the charges on plaquette
A (B). We only need this for small $\chi$, so
\begin{eqnarray}
V_{AB} &=&
9 \Lambda(\phi_A, \phi_B, \Psi) Q_A \chi_A Q_B \chi_B r^6/R^7 \ ,
\nonumber \\ &=&
\frac{9}{4R^7} m_A m_B r^4 \Lambda (\phi_A, \phi_B, \Psi)\ ,
\end{eqnarray}
where $m_i= 2Q_ir \chi_i$ is the $i$th magnetic moment.
For the present situation it seems reasonable to define the
toroidicity $T_i$ of the $i$th plaquette such that 
\begin{eqnarray}
V_{AB} &=& \Lambda(\phi_A, \phi_B, \Psi) T_A T_B / R^7 \ ,
\end{eqnarray}
so that
\begin{eqnarray}
T_i &=& (3/2) m_i r^2 = 3Q_i \chi_i r^3 \ .
\end{eqnarray}
Note that this definition, unlike the standard one,[\onlinecite{TOROID}]
has the property that the magnetostatic interaction energy between two toroids 
is proportional to product of their toroidal `strengths' $T_i$.  To get
$\Lambda (\phi_A, \phi_B, \Psi)$ we may consider the lowest order terms
in the expansion of $V_{AB}$ in powers of $\chi_A$ and $\chi_B$.  Thus
\begin{eqnarray}
\Lambda (\phi_A, \phi_B, \Psi) &=& \frac{R^7}{9Q_AQ_B} \left. 
\frac{d^2}{d\chi_A d \chi_B}
\frac{d^6}{(6!) dr^6} V \right|_{\chi_A=\chi_B=r=0} \ . 
\label{SINEQ} \end{eqnarray}

The phase space of this model is specified as follows.  Each plaquette
is characterized by its center of mass position ${\bf R}$
inside a two dimensional box in the $x$-$y$ plane with respect to which
the plaquette has mirror symmetry.  The $i$th plaquette can assume
an orientation specified by $\phi_i$ and has a toroidal strength
$T_i$ which is an Ising-like variable: $T_i= \pm (3/2)|m_i|r^2$,
according to which of the two ground states the spins occupy.
We assume an orientationally independent interaction between plaquettes
which for concreteness we take to be the Lennard-Jones potential,
\begin{eqnarray}
W_{AB} &=& 4\epsilon \Biggl( \left[ \frac{\sigma}{R}\right]^{12}
- \left[ \frac{\sigma}{R}\right]^{6} \Biggr) \ ,
\label{LJ} \end{eqnarray}
where $\sigma$ and $\epsilon$ are constants.[\onlinecite{LJ}]
Because the system is two dimensional, the plaquettes can not develop
long range order characteristic of a three dimensional solid.[\onlinecite{MW}]
Instead the system can develop various behaviors intermediate between
a conventional solid and an isotropic liquid.  As shown in
Ref. \onlinecite{DRN} the system exhibits a `solid' phase in which
bond-orientational order is truly long range but position correlations
exhibit power-law decay.  The `solid' melts either directly
or indirectly (via a hexatic phase) into an isotropic liquid phase.
Here we focus on the transition as the temperature is lowered through
the value $T_I$ at which the `solid' phase appears and assume that
$kT_I$ is much larger than the magnetic interactions
between plaquettes.  Even in the `solid' phase, there is no true
long range positional order.  Then the spin correlation
function will involve an average over correlations between a spin
at the origin and spins in a distant plaquette.  Since the average spin
in a plaquette is zero, the spin correlation function, like the
positional correlation function, can not display long range order.
In contrast, because there is long-range bond orientational order,
lowering the temperature can lead to phase transitions
due to the interplaquette dipolar interactions.  It is the purpose
of this paper to analyze the symmetry of the resulting ordered phases.
In a likely scenario we find that the `solid' undergoes
two such phase transitions.  At the
first (higher temperature) transition we find that time reversal
symmetry is broken and at the second spatial inversion symmetry is also
broken. Since the underlying system does not have long-range positional
ordering, the spin correlation function itself is never long ranged.
To substantiate this picture it is necessary to analyze the interplaquette
interactions and thereby verify that they lead to Ising-like transitions.

\subsection{Interplaquette Interaction}

To calculate the interaction between two plaquettes, each of which 
is confined to the $x$-$y$ plane, we assume (see Fig. \ref{FIG2})
that plaquette A has charges $\sigma_A Q_A$, where $\sigma_A= \pm 1$,
at (positions relative to the center of the plaquette) 
\begin{eqnarray}
{\bf r}(n_A,\sigma_A) &=& [x(n_A,\sigma_A),y(n_A,\sigma_A)]
\end{eqnarray}
where
\begin{eqnarray}
x(n_A,\sigma_A)
&=& r \cos (\sigma_A \chi_A + 2n_A \pi /3 + \phi_A) \nonumber \\
&=& [r \cos (\chi_A) \cos(2n_A \pi /3 + \phi_A) \nonumber \\
&& - \sigma_A r \sin(\chi_A) \sin(2n_A \pi /3 + \phi_A),
\end{eqnarray}
\begin{eqnarray}
y(n_A, \sigma_A) &=& r \sin (\sigma_A \chi_A+ 2n_A \pi /3+ \phi_A)]
\nonumber \\ &=&
\sigma_A r \sin (\chi_A) \cos (2n_A \pi /3 + \phi_A)
\nonumber \\ && + r \cos (\chi_A) \sin(2 n_A \pi /3 + \phi_A ) ]
\end{eqnarray}
for $n_A=1,2,3$  and similarly for plaquette B.

Then the interaction energy between the two plaquettes is
\begin{eqnarray}
V_{AB} &=& Q_A Q_B  \sum_{n_A,n_B=1}^3
\sum_{\sigma_A \sigma_B = \pm1} \sigma_A \sigma_B
\Biggl[ R^2 \nonumber \\ && +2R \xi\left[ x(n_A,\sigma_A)
- x(n_B,\sigma_B) \right] \nonumber \\ &&
+2R \eta \left[ y(n_A,\sigma_A) - y(n_B,\sigma_B) \right]
\nonumber \\ &&  - 2
{\bf r}(n_A,\sigma_A) \cdot {\bf r}(n_B,\sigma_B) +2r^2 \Biggr]^{-1/2} \ ,
\end{eqnarray}
where $\xi \equiv \cos(\Psi)\equiv X/R$, and $\eta \equiv \sin \Psi \equiv Y/R$.
We drop the last term in $r^2$ since it can not contribute to a term
involving $\sin(3 \chi_A) \sin(3 \chi_B)$.
We want to expand this in powers of $z$ and now we work to first order
in $\chi_A$ and $\chi_B$ by setting $\sin(\chi_X) =\chi_X$ and
$\cos(\chi_X)=1$, where $X$ is either $A$ or $B$.  Thus
\begin{eqnarray}
V_{AB} &=& \frac{Q_AQ_B}{R} \sum_{n_A,n_B=1}^3 \sum_{\sigma_A \sigma_B = \pm 1}
\sigma_A \sigma_B {\cal R}^{-1/2} \ ,
\end{eqnarray}
where
\begin{eqnarray}
&{\cal R}&= 1 + 2 \xi z \Biggl( \cos(2n_A\pi/3 + \phi_A)
- \cos (2n_B\pi/3+\phi_B) \nonumber \\ &-&
\sigma_A \chi_A \sin(2n_A \pi/3+ \phi_A)
+ \sigma_B \chi_B \sin (2n_B \pi/3 + \phi_B) \Biggr)
\nonumber \\
&+& 2\eta z \Biggl( \sigma_A \chi_A \cos(2n_A \pi/3+\phi_A)
\nonumber \\ &-& \sigma_B \chi_B \cos(2n_B \pi /3 + \phi_B) \nonumber \\ 
&& + \sin (2n_A \pi /3 + \phi_A)
- \sin(2n_B \pi /3 + \phi_B ) \Biggr) \nonumber \\
&-& 2z^2 \Biggl( \cos (2n_A \pi /3 + \phi_A)
- \sigma_A \chi_A \sin(2n_A \pi /3 + \phi_A) \Biggr) 
\nonumber \\ && \times
\Biggl( \cos (2n_B \pi /3 + \phi_B) - \sigma_B \chi_B
\sin(2n_B \pi /3 + \phi_B) \Biggr) \nonumber \\ &-& 2z^2 \Biggl(
\sigma_A \chi_A \cos (2n_A \pi /3 + \phi_A) 
+ \sin(2n_A \pi /3 + \phi_A) \Biggr)
\nonumber \\ &\times& \Biggl( \sigma_B \chi_B
\cos (2n_B \pi /3 + \phi_B)
+ \sin(2n_B \pi /3 + \phi_B) \Biggr) \ .
\end{eqnarray}

We now analyze how $V_{AB}$ depends on $\Psi$, i. e. how it depends
on $\xi$ and $\eta$.  Note that $\xi$ and $\eta$ each carry a factor of $z$.
However, factors that do not depend on $\eta$ and $\xi$ carry factors of
$1$ or $z^2$, so that can be no terms at order $z^6$ which are cubic in
$\eta$ or $\xi$.  Therefore $V_{AB}$ can only depend on $\Psi$ through
the argument $6\Psi$.  In addition, for $V_{AB}$ to be invariant under a
global rotation, it can only depend on differences in angle. Also
the result can only be a function of $3\phi_A$ and $3\phi_B$ because it
must be invariant under $\phi_i \rightarrow \phi_i + 2 \pi /3$.  Thus
$\Lambda$ must be of the form
\begin{eqnarray}
&& \Lambda (\phi_A, \phi_B, \Psi)
= a + b \cos(6 \Psi-3 \phi_A - 3 \phi_B) \nonumber \\
&&  + c \cos(3 \phi_A- 3 \phi_B)
+ d \sin(6 \Psi-3 \phi_A - 3 \phi_B)\nonumber \\ && + 
e \sin(3 \phi_A - 3 \phi_B) \ .
\end{eqnarray}
One can view the system looking either along the positive $z$ axis or along
the negative $z$ axis.  Comparing these two views, one sees that a positive
rotation is equivalent to changing the sign of the charges followed by a
negative rotation.  But the interaction energy is invariant under charge 
conjugation.  So
\begin{eqnarray}
\Lambda (\phi_A, \phi_B, \Psi) &=&
a + b \cos(6 \Psi-3 \phi_A - 3 \phi_B) \nonumber \\ &&
+ c \cos(3 \phi_A- 3 \phi_B) \ .
\label{EQLAMB} \end{eqnarray}
Furthermore, consider what happens if we average the interaction over
the orientation $\phi_i$ of one of the plaquettes.  This superposition
of charges leads to a uniformly charge-neutral plaquette.  This 
argument tells us that $a$ in Eq. (\ref{EQLAMB}) must be zero.
To identify the coefficients $b$ and $c$ it suffices to evaluate
\begin{eqnarray}
\Lambda(0,0,\Psi) &=& b \cos (6 \Psi) + c \ .
\label{EVALEQ} \end{eqnarray}

The explicit evaluation of $\Lambda(0,0,\Psi)$ is carried out in the
Appendix.  Then use of Eq. (\ref{EQLAMB}) implies the result
\begin{eqnarray}
\Lambda &=& \frac{10,395}{32} \cos(6 \Psi- 3 \phi_A - 3 \phi_B)
\nonumber \\ && -\frac{225}{32} \cos(3 \phi_A - 3 \phi_B) \ .
\label{RESEQ} \end{eqnarray}

\begin{figure}[ht]
\begin{center}
\includegraphics[width=8.0 cm]{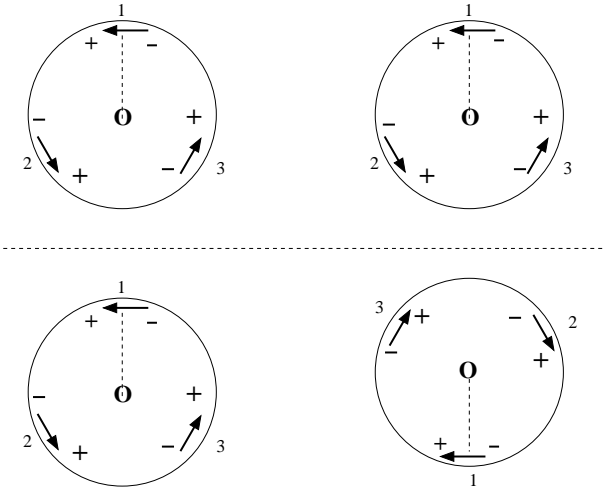}
\caption{\label{FIG3} Top: two interacting plaquettes (both with
$\phi=\pi /2$ and $Q=+1$).  Bottom: same as the top configuration
except that the right-hand plaquette is rotated through an angle
$\Delta \phi=\pi$ and sign of the charge is now $Q=-1$.
These two configurations have the same interaction energy truncated at
order $R^{-7}$.  This is the gauge invariance of Eq. (\ref{QPHIEQ}).}
\end{center}
\end{figure}

It is important to note a local symmetry.  With the interactions so far
postulated, the Hamiltonian is invariant under the local transformation
\begin{eqnarray}
Q_i \rightarrow -Q_i \ , \hspace{1 in}
\phi_i \rightarrow \phi_i + \pi \ .
\label{QPHIEQ} \end{eqnarray}
This is a nontrivial symmetry that indicates that the two configurations
shown in Fig. \ref{FIG3} have the same energy at leading order in the
multipole expansion.  Note that changing the sign of
$Q_i$ is equivalent to changing the sign of the toroidal moment $T_i$.
As a result of this local gauge symmetry it follows from
Elitzur's theorem[\onlinecite{ELITZ}] that even though there is
long range order in the variable $Q \sin(3\phi)$, there is no long range
order in either $Q$ or $\sin(3 \phi)$.  

\section{PHASE TRANSITIONS OF THIS MODEL}

Here we give a more detailed analysis of the phase transitions
within this model.  We assume that the magnetic anisotropy
energy that aligns the spins along a fixed direction in each plaquette
and the isotropic interactions of Eq. (\ref{LJ}) between plaquettes
are dominant.  These assumed interactions do not depend on the
orientation of either of the interacting plaquettes.
(With this assumption Elitzur's theorem applies.)  Accordingly,
as the temperature is lowered, this two dimensional system will
undergo a phase transition at a temperature $T_I$ into a `solid' phase
with long-range bond-orientational order, but {\it no} long range
positional order.]\onlinecite{DRN}] It is obvious that in this phase
both spatial inversion and time reversal symmetries are maintained.
 
\begin{figure}[ht]
\begin{center}
\includegraphics[width=8.0 cm]{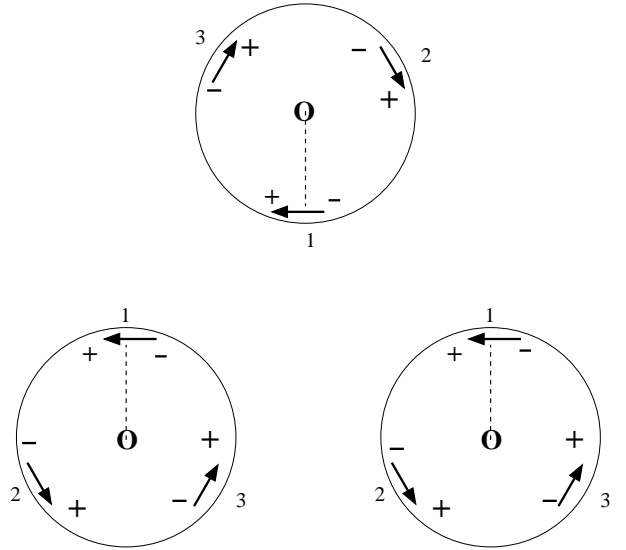}
\caption{\label{FIG4} Long range order with unbroken gauge symmetry.
Note that the top-most plaquette is in the gauge transformed state
according to Eq. (\ref{QPHIEQ}).  Here we illustrate the case when the
order parameter $Q\sin(3 \phi )$ is negative.}
\end{center}
\end{figure}

As one further reduces the temperature, the interplaquette dipolar
interactions come into play and can cause order to develop
consistent with the local gauge symmetry.  To see what sort of
order develops we introduce the appropriate gauge invariant variables
\begin{eqnarray}
X_i &=& T_i \cos(3 \phi_i) \ , \ \ \ \
Y_i = T_i \sin(3 \phi_i) \ .
\end{eqnarray}
Because bond orientational order is maintained, we can treat each
molecule as being surrounded by a hexagon of neighboring plaquettes
and the orientation of this hexagon of neighbors is maintained over
the entire system.  Accordingly, we can define $\Psi$ as being
measured relative to the direction between the central plaquette and
one of its neighbors.  So we may take $6 \Psi /(2 \pi)$ to be
an integer for all nearest neighbor interactions.  We do not consider
further neighbor interactions in view of how rapidly the interplaquette
interaction falls off with separation.  For simplicity we work as
if we have a two dimensional triangular lattice. Thus we analyze
the model with orientationally-dependent interactions 
\begin{eqnarray}
V_{AB} &=&  \frac{10,170T_AT_B}{32R^7}
\Biggl[  \cos(3\phi_A) \cos(3\phi_B) \nonumber \\ &&
- \frac{10,620T_AT_B}{32R^7} \sin (3\phi_A) \sin(3\phi_B) \Biggr] \ .
\end{eqnarray}
So we have a two dimensional anisotropic rotor model which is in the
same universality class as the two dimensional Ising model.  Such
models have been widely studied.[\onlinecite{ROTOR,ABH1}] To analyze the
phase transitions within this model we invoke mean-field 
theory, within which the Landau free energy,
${\cal F}$, in terms of the Fourier transforms of $X_i$ and $Y_i$
(temporarily assuming a triangular lattice of lattice constant $a$)
assumes the form
\begin{eqnarray}
{\cal F} &=& \frac{1}{2} \sum_{\bf q} \Biggl( [ckT + \mu({\bf q})]X({\bf q})
X(({\bf q})^* \nonumber \\ &&
+ [ckT + \nu({\bf q})] Y({\bf q}) Y(({\bf q})^* \Biggr) \ ,
\end{eqnarray}
at quadratic order, where $c$ is a constant of order unity, and
$\mu ({\bf q})$ and $\nu ({\bf q})$ are the
Fourier transforms of the potential:
\begin{eqnarray}
\mu ({\bf q}) &=& 2A [\cos(aq_x) + 2 \cos(aq_x/2) \cos(\sqrt 3 a q_y/2) \ ]
\nonumber \\
\nu ({\bf q}) &=& 2B [\cos(aq_x) + 2 \cos(aq_x/2) \cos(\sqrt 3 a q_y/2) \ ] \ ,
\end{eqnarray}
where
\begin{eqnarray}
A &=& \frac{10,170T^2}{32R^7} \ , \ \ \ \
B = - \frac{10,620T^2}{32R^7} \ ,
\end{eqnarray}
where $T=(3/2) m r^2$. Thus as the temperature is lowered the system will
develop long range ``ferro" order (order at zero wave vector)
in the variable $Y \equiv T \sin(3\phi)$, but not, as was said, in either
$T$ or $\sin(3 \phi)$ separately.  This type of order is illustrated
in Fig. \ref{FIG4}.  This transition occurs at a temperature of order
$kT_{II}=6|B| \approx 2000T^2 /R^7$. 

\begin{figure}[ht]
\begin{center}
\includegraphics[width=6.0 cm]{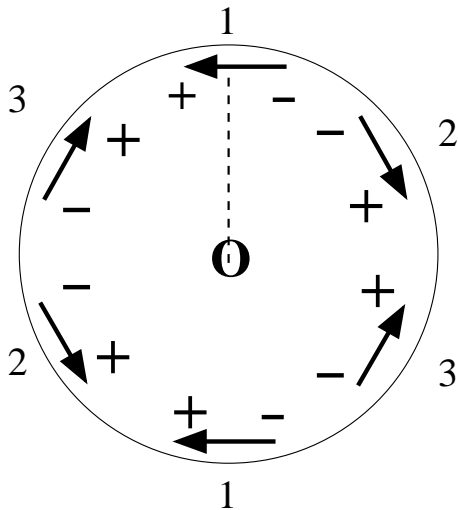}
\caption{\label{FIG5} The ordered component of the density
matrix incorporating states S' and S''.  This state is not
time reversal invariant but is inversion symmetric because
spins being pseudo vectors do not change their orientations
under inversion.}
\end{center}
\end{figure}

We now discuss whether time reversal (T) symmetry or spatial inversion
(P) symmetry is broken in this phase.  As a preliminary, note that
the single particle density matrix assigns the probabilities
$p/2$, $p/2$, and $1-p$ to the states S, S', and S'', respectively, 
where S is the state of the top-most plaquette in Fig. \ref{FIG4},
S' is the state of a plaquette in the bottom row of Fig. \ref{FIG4},
and S'' is the completely disordered state. The interpretation of this
density matrix is shown in Fig. \ref{FIG5}, where we see that time
reversal symmetry is broken but inversion symmetry is maintained.
(If the dipole were electric dipoles, then inversion symmetry would be
broken.)

\begin{figure}[ht]
\begin{center}
\includegraphics[width=8.5 cm]{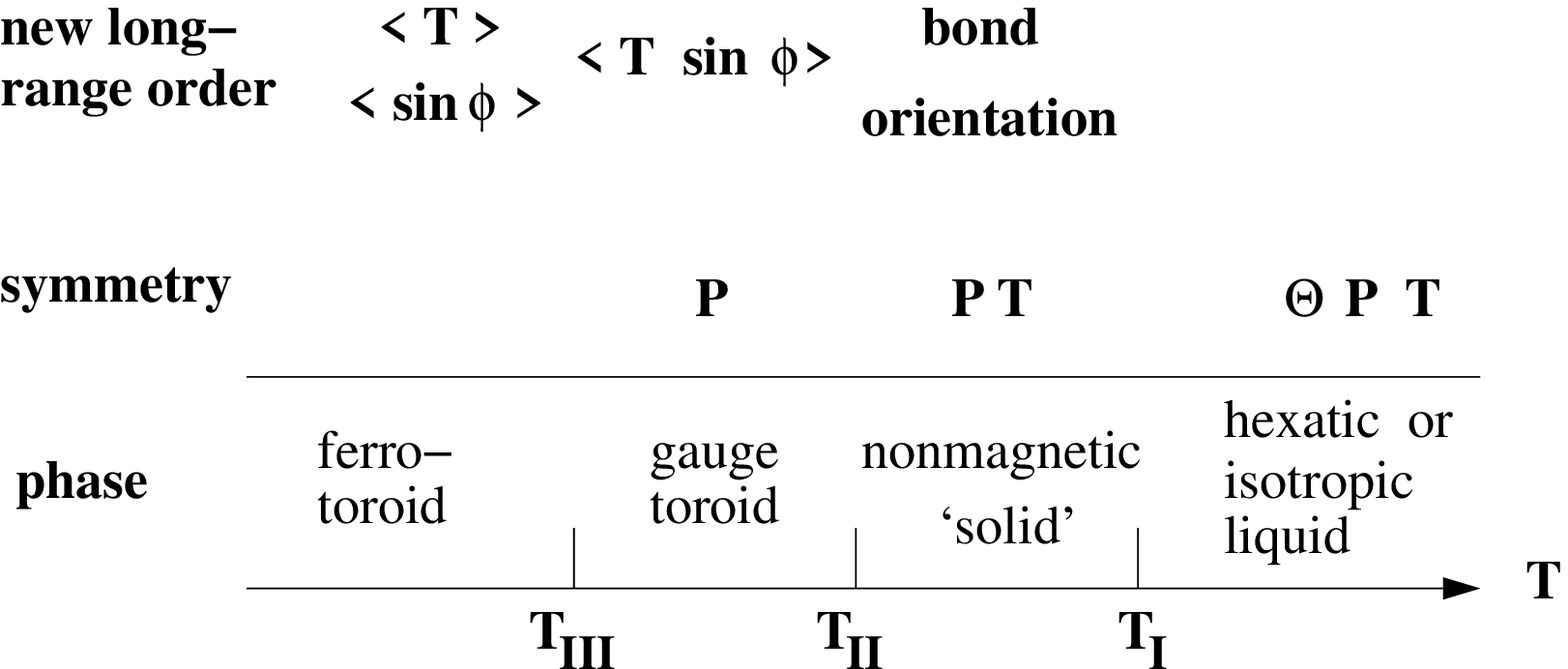}
\caption{\label{FIG6} Symmetry of the various phases. Here the symmetries
are $T=$ time reversal symmetry, $P=$ spatial inversion symmetry,  and
$\Theta=$ continuous rotational symmetry.}
\end{center}
\end{figure}

In this phase (which we call the `gauge toroid' phase), because
of the gauge symmetry, there is no long range order in the orientation 
of the plaquettes.  This is a result of our assumption that the
interaction between plaquettes is independent of their orientations.
Of course, this assumption is not 
consistent with the three-fold symmetry implied by the existence of
the spins.  Accordingly, we now take proper account of this three-fold
symmetry by introducing small bulges in the plaquettes at the locations
of the three spins.  This will give rise to an interplaquette interaction
which breaks the local gauge symmetry.  For simplicity we take
this gauge symmetry breaking interaction to be of the form
\begin{eqnarray}
V' &=& v \cos(6 \Psi - 3 \phi_A - 3 \phi_B) \ . 
\end{eqnarray}
Now we again set $6\Psi/(2 \pi)$ to be an integer and we only need
to consider interactions involving $\sin(3 \phi )$, so effectively
\begin{eqnarray}
V' &=& -v \sin(3 \phi_A) \sin(3 \phi_B) \ .
\end{eqnarray}
Now what happens depends on the sign of $v$.  If $v$ is 
positive, then we have a ferro arrangement of plaquettes, so that
all plaquettes are in the same state (either as those in the bottom
row of Fig. \ref{FIG4} or as that in the top  row of Fig. \ref{FIG4}).
Because toroidicity and orientation are strongly
coupled, this state is ferrotoroidal.  If $v$ is negative, then we have an
antiferro arrangement of plaquettes into the so-called `root-3'
structure discussed recently in connection
with charge ordering in lutetium ferrite.[\onlinecite{ABH2}]
In this state we have antiferrotoroidicity.
Here we are mainly interested in displaying a ferro state, so
we take $v$  to be positive.  This final ordering transition
which breaks local gauge symmetry will occur at a temperature
of order $kT_{III}=6v$, which we assume to be much smaller than
$kT_{II}$.  This transition is also in the same universality
class as the two dimensional Ising model. At this transition
spatial inversion symmetry is broken. 
The symmetry of the various phases is summarized in Fig. \ref{FIG6}.

In Ref. \onlinecite{NAS2} toroidicity has been discussed in connection
with the magnetoelectric effect.  However, instead of using the symmetry
of the crystal (see Ref. \onlinecite{HS}), they used the symmetry of
free space to obtain a simplified relation between the toroidicity and
the linear magnetoelectric tensor.  Here we invoke the symmetry of the
two dimensional system to write the magnetoelectric free energy as
a function of the electric field ${\bf E}$ and the magnetic field
${\bf H}$ as
\begin{eqnarray}
{\cal F}_{\rm ME} &=& T_z \Biggl( A [H_xE_y-H_yE_x] + BH_zE_z 
\nonumber \\ && \ + C[H_xE_x+H_yE_y] \Biggr) \ ,
\end{eqnarray}
where $T_z$ is the $z$-component of the toroidicity, ${\bf T}$.
We may check that this form is consistent with the symmetry of the system,
keeping in mind that spin and the magnetic field are both pseudo vectors
but ${\bf T}$ is a real vector. Accordingly, both $T_z$ and the factor in the
large brackets are odd under the mirror $z \rightarrow -z$.  Also
all the terms are invariant under a rotation about the $z$ axis.
Thus 
\begin{eqnarray}
- \frac{\partial^2 {\cal F}}{\partial H_x \partial E_y} &=&
\frac{\partial M_x}{\partial E_y} = \frac{\partial P_y}{\partial H_x} = -A
\nonumber \\
- \frac{\partial^2 {\cal F}}{\partial H_y \partial E_x} &=&
\frac{\partial M_y}{\partial E_x} = \frac{\partial P_x}{\partial H_y} = A
\nonumber \\
- \frac{\partial^2 {\cal F}}{\partial H_z \partial E_z} &=&
\frac{\partial M_z}{\partial E_z} = \frac{\partial P_z}{\partial H_z} = -B
\nonumber \\
- \frac{\partial^2 {\cal F}}{\partial H_x \partial E_x} &=&
\frac{\partial M_x}{\partial E_x} = \frac{\partial P_x}{\partial H_x} = -C
\nonumber \\
- \frac{\partial^2 {\cal F}}{\partial H_y \partial E_y} &=&
\frac{\partial M_y}{\partial E_y} = \frac{\partial P_y}{\partial H_y} = -C \ .
\end{eqnarray}
These elements of the linear magnetoelectric tensor are only nonzero
in the ferrotoroidal phase.  The combination of the three-fold axis and
the $x$-$y$ reflection plane guarantee that the spontaneous polarization
is zero in all these phases. (If the discs were asymmetric with
respect to this mirror, then a spontaneous polarization along $z$
would be allowed in the ferrotoroidal phase.\cite{KAPLAN})

\section{DISCUSSION AND CONCLUSION}

The nonmagnetic `solid'
to gauge toroid transition at $T_{II}$ was analyzed assuming
no gauge breaking interactions. We argue that the introduction of  the
small gauge breaking interaction will not qualitatively modify the phase
diagram of Fig. \ref{FIG6}, because it will take a finite interaction to
orientationally order the plaquettes. Similarly, including higher order
terms in the multipole expansion will not alter our conclusions as long
as $z \equiv r/R \ll 1$.  One possible difficulty in constructing the
system of discs analyzed in this paper is that it may
be difficult to achieve equilibrium statistics within the manifold
of the dipolar spin ground states.  But perhaps it is not crucial that
the gap between the two spin ground states and the qualitatively
different excited  spin states is very large. It would, of course,
be extremely interesting to observe the linear magnetoelectric effect in
a system such as this.

\vspace{0.3 in}

\centerline{{\bf ACKNOWLEDGMENTS}}

\vspace{0.2 in}

I would like to thank N. A. Spaldin for introducing me to this subject
and C. Kane for helpful discussions.  I am also very grateful
to Michael Cohen for corroborating Eq. (\ref{RESEQ}).

\begin{appendix}
\section{Evaluation of $\Lambda(\phi_A, \phi_B, \Psi)$}

Write
\begin{eqnarray}
V_{AB} &=& \frac{Q_AQ_B}{R} \sum_{n_A,n_B=1}^3 \sum_{\sigma_A \sigma_B}
\sigma_A \sigma_B \Biggl[ 1 + Az + B_1z \sigma_A \chi_A
\nonumber \\ && +  B_2 z \sigma_B \chi_B + C_1 z^2 \sigma_A \chi_A
+ C_2 z^2 \sigma_B \chi_B \nonumber \\ &&
+ Dz^2 \sigma_A \sigma_B \chi_A \chi_B + Ez^2 \Biggr]^{-1/2} \ .
\end{eqnarray}
Here
\begin{eqnarray}
A &=& 2 \xi \left[ \cos (2n_A \pi /3 + \phi_A)
- \cos (2n_B \pi /3 + \phi_B) \right]
\nonumber \\ && + 2 \eta \left[ \sin (2n_A \pi /3 + \phi_A) 
- \sin (2n_B \pi /3+\phi_B) \right] \nonumber \\
B_1 &=& - 2 \xi \sin (2n_A \pi /3+\phi_A)
+ 2 \eta  \cos (2n_A \pi /3+\phi_A) \nonumber \\
B_2 &=& 2 \xi \sin (2n_B \pi /3+\phi_B)
- 2 \eta  \cos (2n_B \pi /3+\phi_B)
\nonumber \\
C_1 &=& 2 \sin (2n_A \pi /3+\phi_A) \cos(2n_B \pi /3+\phi_B) 
\nonumber \\ && - 2 \cos (2n_A \pi /3+\phi_A) \sin(2n_B \pi /3+\phi_B)
\nonumber \\
C_2 &=& 2 \cos (2n_A \pi /3+\phi_A) \sin(2n_B \pi /3+\phi_B)
\nonumber \\ && - 2 \sin (2n_A \pi /3+\phi_A) \cos(2n_B \pi /3+\phi_B)
= - C_1 \nonumber \\
D &=& -2  \sin (2n_A \pi /3+\phi_A) \sin(2n_B \pi /3+\phi_B)
\nonumber \\ && - 2 \cos(2n_A \pi /3+\phi_A) \cos(2n_B \pi /3+\phi_B)
\nonumber \\
E &=& -2  \cos (2n_A \pi /3+\phi_A) \cos(2n_B \pi /3+\phi_B)
\nonumber \\ && - 2 \sin(2n_A \pi /3+\phi_A) \sin(2n_B \pi /3+\phi_B) = D \ .
\nonumber \\ &&
\end{eqnarray}
Thus, when the sums over $\sigma_A$ and $\sigma_B$ are performed, we have
\begin{eqnarray}
V_{AB} &=& \frac{4Q_AQ_B}{R} \sum_{n_A,n_B=1}^3
\nonumber \\ && \times \Biggl[ 1 + Az + Dz^2 \Biggr]^{-1/2}
\Biggl[  - \frac{D z^2 \chi_A \chi_B}{2(1+Az+Dz^2)} \nonumber \\
&+& \frac{3[B_1B_2z^2 + (B_2-B_1)C_1z^3 -C_1^2z^4]\chi_A \chi_B }
{4(1+Az+Dz^2)^2} \Biggr] \ .
\nonumber \\ && 
\end{eqnarray}
We have to evaluate this at order $z^6$. According to Eq. (\ref{SINEQ})
\begin{eqnarray}
&& \Lambda (\phi_A, \phi_B, \Psi) = \frac{4}{9(6!)}
\left. \frac{d^6}{dz^6} \right|_{z=0} \sum_{n_A,n_B=1}^3 
\nonumber \\ && \times \Biggl[ - \frac{D z^2 }{2(1+Az+Dz^2)^{3/2}}
\nonumber \\ &&
+ \frac{3[B_1B_2z^2+(B_2-B_1) C_1z^3 - C_1^2 z^4]}{4(1+Az+Dz^2)^{5/2}}
\Biggr] \ .
\end{eqnarray}
Thus
\begin{eqnarray}
&& \Lambda  = \frac{4}{9} \sum_{n_A,n_B=1}^3 \Biggl\{ 
- \frac{(-3/2!)D^3}{2(-7/2!)(2!)}
- \frac{(-3/2!)D^2A^2}{2(-9/2!)(1!)(2!)} \nonumber \\
&-& \frac{(-3/2!)DA^4}{2(-11/2!)(0!)(4!)}
+ \frac{3(-5/2!) B_1B_2D^2}{4(-9/2!)(2!)}
\nonumber \\ &&
+ \frac{3(-5/2!) B_1B_2DA^2}{4(-11/2!)(1!)(2!)}
+ \frac{3(-5/2!) B_1B_2A^4}{4(-13/2!)(0!)(4!)}
\nonumber \\
&& + \frac{3(-5/2!) (B_2-B_1)C_1DA}{4(-9/2!)(1!)(1!)}
\nonumber \\ &&
+ \frac{3(-5/2!) (B_2-B_1)C_1A^3}{4(-11/2!)(0!)(3!)}
\nonumber \\
&& - \frac{3(-5/2!) C_1^2D}{4(-7/2!)(1!)}
- \frac{3(-5/2!) C_1^2A^2}{4(-9/2!)(0!)(2!)} \Biggr\}
\nonumber \\ &\equiv& \frac{4}{9} \sum_{n=1}^{12} \Lambda_n  \  ,
\label{EQLAM} \end{eqnarray}
where we number the terms with $(B_2-B_1)$ separately.

To implement Eq. (\ref{EVALEQ}) we write
\begin{eqnarray}
\Lambda(0,0, \Psi) &=& (4/9) \left[ \Lambda_6(0,0,\Psi) +
\Delta \Lambda(0, 0, \Psi) \right] \ ,
\end{eqnarray}
where $\Lambda_6$ is the term in $\Lambda$ proportional to
$B_1B_2A^4$ and $\Delta \Lambda$ contains the remaining terms written
in Eq. (\ref{EQLAM}). Note that $\Lambda_6$ is the only term
which is sixth order in $\xi$ and $\eta$ and is therefore the only
term which can contribute to the term in Eq. (\ref{EQLAMB})
involving $6\Psi$. Thus we write
\begin{eqnarray}
\Lambda_6 &=& \alpha + \beta \cos(6\Psi-3 \phi_A - 3 \phi_B) 
\nonumber \\ && + c_6 \cos(3\phi_A- 3\phi_B)
\end{eqnarray}
and $\Delta \Lambda$ is a constant which we can evalute by setting
$\Psi=0$.
\begin{eqnarray}
\Delta \Lambda = a' + c' \cos(3 \phi_A-3 \phi_B) \ .
\end{eqnarray}
We first evaluate $\Lambda_6$ for $\phi_A=\phi_B=0$ with
$\Psi$ arbitrary. It is convenient to introduce the
notation $(x,y)$, where $x$ and $y$ assume the values 1, $c$, or $s$,
for unity, the cosine function and the sine function. The first
argument is that for plaquette A and  the second is that for plaquette
B.  Thus $(c,cs)\equiv \cos(2 n_A \pi /3) \cos(2 n_B \pi /3)
\sin(2 n_B \pi /3)$ and $(1,c^2)\equiv \cos^2(2n_B \pi /3 )$. 
We have the sums over $n$:
\begin{eqnarray}
&& \sum_n (\cos(2 n \pi /3) = \sum_n \sin(2 n \pi /3) \cos^k (2 n \pi /3) = 0
\nonumber \\
&& \sum_n \sin^2 (2 n \pi /3) = \sum_n \cos^2 (2 n \pi /3) = 3/2
\nonumber \\
&& \sum_n \cos^3(2n \pi /3) = - \sum_n \sin^2(2 n \pi /3) \cos(2n \pi /3)
= 3/4
\nonumber \\
&& \sum_n \cos^2 (2 n \pi /3) \sin^2(2 n \pi /3) =  3/8
\nonumber \\
&& \sum_n \cos^4(2n \pi /3) = \sum_n \sin^4 (2 n \pi /3) = 9/8 \ .
\end{eqnarray}

With these understandings (and with the sums over $n_A$ and $n_B$
implied) we write 
\begin{eqnarray}
{\cal S}_6 &=& \sum_{n_A,n_B=1}^3 B_1B_2 A^4 \ , 
\end{eqnarray}
so that
\begin{eqnarray}
{\cal S}_6 &=& -64 [\xi(s,1) - \eta(c,1)][\xi(1,s)-\eta(1,c)] \nonumber \\
&& \times \Biggl( \xi[(c,1)-(1,c)] + \eta[(s,1)-(1,s)] \Biggr)^4
\nonumber \\ &=&
-64 [ \xi^2(s,s) + \eta^2 (c,c) -\xi\eta(s,c) - \xi\eta(c,s)]
\nonumber \\ && \times
\Biggl( \xi[(c,1)-(1,c)] + \eta[(s,1)-(1,s)] \Biggr)^4 \nonumber \\ &=&
-64 [\xi^2 (s,s)] \Biggl( \eta^4 [ -4 (s^3,s)-4(s,s^3)]
\nonumber \\ &&  - 12 \eta^2 \xi^2 (s,s) [(c,1)-(1,c)]^2\Biggr)
\nonumber \\ && - 64 [ \eta^2 (c,c)]
\Biggl( \xi^4 [(c,1)-(1,c)]^4 \nonumber \\ &&
+ 6 \xi^2 \eta^2 [(c,1)-(1,c)]^2 [
(s^2,1) + (1,s^2)]
\nonumber \\ &&  + \eta^4 [(s,1)-(1,s)]^4 \Biggr) \nonumber \\ &&
\nonumber \\ &&
+ 64 \xi \eta (s,c) \Biggl( 4 \xi^3 \eta [(c,1)-1,c)]^3 (s,1) 
\nonumber \\ && + 4 \eta^3 \xi
[(c,1)-(1,c)][(s^3,1) + 3(s,s^2)] \Biggr) \nonumber \\ &&
+ 64 \xi \eta (c,s) \Biggl( - 4 \xi^3 \eta [(c,1)-(1,c)]^3 (1,s)
\nonumber \\ &&
- 4 \eta^3 \xi[(c,1)-(1,c)][(1,s^3)+3(s^2,s)] \Biggr) \nonumber \\ &=&
\eta^6 \Biggl( -64 (c,c)[(s,1)-(1,s)]^4 \Biggr) \nonumber \\ &&
+ \eta^4 \xi^2 \Biggl( 256 [(s^4,s^2)+(s^2,s^4)]
\nonumber \\ && 
- 384 (c,c)[(c,1)-(1,c)]^2 [(s^2,1)+ (1,s^2)] \nonumber \\ &&
+ 256 [(s,c)+(cs)] [(c,1)-(1,c)][(s^3,1) + 3 (s,s^2)] \Biggr) \nonumber \\ &&
+ \eta^2 \xi^4 \Biggl( 768(s^2, s^2) [(c,1)-(1,c)]^2 
\nonumber \\ && -64 (c,c) [(c,1)-(1,c)]^4 \nonumber \\ && 
+ 256 [(s^2,c)+(c,s^2)] [(c,1)-(1,c)]^3 \Biggr) \ .
\end{eqnarray}
When simplified this is
\begin{eqnarray}
{\cal S}_6 &=&
\eta^6 [-64(6)(cs^2,cs^2)] + 512 \eta^4 \xi^2 (s^4,s^2)
\nonumber \\ &&
-768 \eta^4 \xi^2 [-2(c^2s^2,c^2)+(cs^2,c^3)] \nonumber \\ &&
+512 \eta^4 \xi^2 [ - (s^4,c^2) + 3(s^2c,s^2c)-3(s^2,s^2c^2)]
\nonumber \\ &&
+768 \eta^2 \xi^4 [(s^2c^2,s^2) - 2(s^2c,s^2c) + (s^2,s^2c^2)]
\nonumber \\ &&
-64 \eta^2 \xi^4 [-4 (c^4,c^2) + 6(c^3,c^3) -4(c^2,c^4)]
\nonumber \\ &&
+512 \eta^2 \xi^4 [-3(s^2c^2,c^2) + 3(s^2c,c^3) -(s^2,c^4) \nonumber \\
&=& \eta^6[-384] \left[ - \frac{3}{4} \right]^2
+ 512 \eta^4 \xi^2 \left[ \frac{9}{8} \right] \left[ \frac{3}{2} \right]
\nonumber \\ && -768 \eta^4 \xi^2
\Biggl( - 2 \left[\frac{3}{8} \right] \left[ \frac{3}{2} \right] + \left[
- \frac{3}{4} \right] \left[ \frac{3}{4} \right] \Biggr) \nonumber \\
&+& 512 \eta^4 \xi^2 \Biggl( - \left[ \frac{9}{8} \right] \left[ \frac{3}{2}
\right] + 3 \left[ - \frac{3}{4} \right]^2 - 3 \left[ \frac{3}{2} \right]
\left[ \frac{3}{8} \right] \Biggr) \nonumber \\
&+& 768 \eta^2 \xi^4 \Biggl( \left[ \frac{3}{8} \right] \left[ \frac{3}{2} 
\right] - 2 \left[ - \frac{3}{4}\right]^2 + \left[ \frac{3}{2} \right]
\left[ \frac{3}{8} \right] \Biggr) \nonumber \\
&-& 64 \eta^2 \xi^4 \Biggl( -4 \cdot 2 \left[ \frac{9}{8} \right]
\left[ \frac{3}{2} \right] + 6 \left[ \frac{3}{4} \right]^2 \Biggr)
\nonumber \\ &+&
512 \eta^2 \xi^4 \Biggl( -3 \left[ \frac{3}{8} \right] \left[ \frac{3}{2}
\right] + 3 \left[
- \frac{3}{4} \right] \left[ \frac{3}{4} \right] - \left[ \frac{3}{2} \right]
\left[ \frac{9}{8} \right] \Biggr) \nonumber \\
&=& - 216 \eta^6 + \eta^4 \xi^2 [1296] - \eta^2 \xi^4 [1944] \ ,
\end{eqnarray}
This is
\begin{eqnarray}
{\cal S}_6 &=& \Biggl( -108 (\xi^2+\eta^2)^3 \nonumber \\ &&
-108 (\eta^6 -15 \eta^4\xi^2
+15 \eta^2 \xi^4 - \xi^6) \Biggr) \nonumber \\
&=& 108 [ \cos(6\Psi) -1] \ .
\end{eqnarray}
This leads to the result
\begin{eqnarray}
\Lambda_6(\phi_A, \phi_B, \Psi)
&=& \frac{93,555}{128} \Biggl[ - \cos(3 \phi_A - 3 \phi_B)
\nonumber \\ &&
+ \cos(6 \Psi - 3 \phi_A - 3 \phi_B) \Biggr] \ .
\end{eqnarray}

We now evaluate the other sums. But since $\Delta \Lambda$ does
not depend on $\Psi$, we set $\xi=1$ and $\eta=0$.  Then we have
(again the sums over $n_A$ and $n_B$ are left implicit)
\begin{eqnarray}
\sum D^3 &\equiv& {\cal S}_1 =  -8 [(s,s) + (c,c)]^3
\nonumber \\ &=& -8[ 3(s^2c,s^2c) + (c^3,c^3)]
\nonumber \\ &=& -8 \Biggl( 3\left[ -\frac{3}{4} \right]^2 +
\left[ \frac{3}{4} \right]^2 \Biggr) 
= - 18 \ .
\end{eqnarray}
If ${\cal S}_2 = \sum D^2A^2$, then, with the sums implicit, we have
\begin{eqnarray}
{\cal S}_2 &=& 16 [(s^2,s^2) + 2(sc,sc) + (c^2,c^2)]
\nonumber \\ && \times [(c,1)-(1,c)]^2
\nonumber \\ &=&
16[(s^2c^2,s^2) + (c^4,c^2) -2 (s^2c,s^2c) -2(c^3,c^3) \nonumber \\ &&
+(s^2,s^2c^2) + (c^2,c^4)] \nonumber \\ &=&
32 \Biggl( \left[ \frac{3}{8} \right] \left[ \frac{3}{2} \right]
+ \left[ \frac{9}{8} \right] \left[ \frac{3}{2} \right] 
- \left[ -\frac{3}{4} \right]^2 - \left[ \frac{3}{4} \right]^2
\Biggr) \nonumber \\ &=& 36 \ .
\end{eqnarray}
If ${\cal S}_3 = \sum DA^4$, then, with the sums implicit, we have
\begin{eqnarray}
{\cal S}_3 &=& 32 [-(ss)-(cc)] [(c,1)-(1,c)]^4
\nonumber \\ &=& 64 [ 2 (c^4,c^2) -3 (c^3,c^3) +2(c^2,c^4)]
\nonumber \\ &=&
64 \Biggl( 4 \left[ \frac{9}{8} \right] \left[
\frac{3}{2} \right] - 3 \left[ \frac{3}{4} \right]^2 \Biggr)
\nonumber \\ &=& 324 \ .
\end{eqnarray}
If ${\cal S}_4 = \sum B_1B_2D^2$, then, with the sums implicit, we have
\begin{eqnarray}
{\cal S}_4 &=& -16 (s,s)[(s,s)+(c,c)]^2 \nonumber \\ &=&
-32 (s^2c,s^2c) = -32 \left[ - \frac{3}{4} \right]^2
= -18 \ .
\end{eqnarray}
If ${\cal S}_5 = \sum B_1B_2DA^2$, then, with the sums implicit, we have
\begin{eqnarray}
{\cal S}_5 &=& 32 (s,s) [(s,s) + (c,c)] [(c,1)-(1,c)]^2
\nonumber \\ &=& 32 (s^2,s^2)[(c^2,1) - 2(c,c)+(1,c^2)]
\nonumber \\ &=& 
32[(s^2c^2,s^2)-2(cs^2,cs^2)+(s^2,s^2c^2)] \nonumber \\ &=& 
64 \Biggl( \left[ \frac{3}{8} \right]
\left[ \frac{3}{2} \right]  - \left[ - \frac{3}{4} \right]^2 \Biggr)
\end{eqnarray}
If ${\cal S}_7 = \sum B_2C_1DA$, then, with the sums implicit, we have
\begin{eqnarray}
{\cal S}_7 &=& -16  (1,s)[(s,c)-(c,s)][(c,c)+(s,s)][(c,1)-(1,c)]
\nonumber \\ &=& 16[-(s,sc)+(c,s^2)] [(c,c)+(s,s)][(c,1)-(1,c)]
\nonumber \\ &=& 16[-(s^2,s^2c)+(c^2,cs^2)] [(c,1)-(1,c)]
\nonumber \\ &=& 16[-(s^2c,s^2c)+(c^3,cs^2)+(s^2,s^2c^2) - (c^2,c^2s^2)]
\nonumber \\ &=& 16
\Biggl(  - \left[- \frac{3}{4} \right]^2
+ \left[ \frac{3}{4} \right] \left[ -\frac{3}{4} \right] \Biggr)
= - 18 \ .
\end{eqnarray}
\begin{eqnarray}
\sum B_1C_1DA &\equiv& {\cal S}_8 = 
-16(s,1)[(s,c)-(c,s)] \nonumber \\ && \times
[(s,s) + (c,c)][(c,1)-(1,c)] \nonumber \\ &=& - \sum B_2C_1DA
\end{eqnarray}
If ${\cal S}_9 = \sum B_2C_1A^3$, then, with the sums implicit, we have
\begin{eqnarray}
{\cal S}_9 &=& 32(1,s)[(s,c)-(c,s)] [(c,1)-(1,c)]^3 
\nonumber \\ &=& -32 (c,s^2)[(c,1)-(1,c)]^3
\nonumber \\ &=& -32[(c^4,s^2)-3(c^3,cs^2)+3(c^2,c^2s^2)]
\nonumber \\ && \nonumber \\ &=&
-32 \Biggl( \left[ \frac{9}{8} \right] \left[ \frac{3}{2} \right] 
- 3 \left[ - \frac{3}{4} \right] \left[ \frac{3}{4} \right] +3 \left[
\frac{3}{2} \right] \left[ \frac{3}{8} \right] \Biggr) 
\nonumber \\ &=& -162 \ .
\end{eqnarray}
\begin{eqnarray}
&& \sum B_1C_1A^3\equiv {\cal S}_{10} = 
32[\xi(s,1)-\eta(c,1)][(s,c)-(c,s)] \nonumber \\ &\times&
\Biggl( \xi[(c,1)+(1,c)] + \eta[(s,1)+(1,s)] \Biggr)^3
\nonumber \\ && = - \sum B_2C_1A^3 
\end{eqnarray}
If ${\cal S}_{11} = \sum C_1^2D$, then, with the sums implicit, we have
\begin{eqnarray}
{\cal S}_{11} &=& -8[(s,c)-(c,s)]^2[(c,c)+(s,s)] \nonumber \\ &=&
-8[(s^2,c^2)+(c^2,s^2)](c,c) + 16(cs,cs)(s,s) \nonumber \\ &=&
-8 \Biggl( 2 \left[-\frac{3}{4}\right] \left[ \frac{3}{4} \right]
- 2 \left[ - \frac{3}{4} \right]^2 \Biggr)
= 18 \ .
\end{eqnarray}
If ${\cal S}_{12} = \sum C_1^2A^2$, then, with the sums implicit, we have
\begin{eqnarray}
{\cal S}_{12} &=& 16 [(s,c)-(c,s)]^2[(c,1)-(1,c)]^2 \nonumber \\
&=& 16[(s^2,c^2)+(c^2,s^2)][(c^2,1)-2(c,c)+(1,c^2)] \nonumber \\
&=& 32 [(s^2c^2,c^2) + (c^4,c^2) - (s^2c,c^3)- (c^3,s^2c)]
\nonumber \\ &=&
32 \Biggl( \left[ \frac{3}{8}\right] \left[ \frac{3}{2} \right]
+ \left[ \frac{3}{2} \right] \left[ \frac{9}{8} \right]  
- 2  \left[ - \frac{3}{4} \right] \left[ \frac{3}{4} \right] \Biggr)
\nonumber \\ &=& 108 \ .
\end{eqnarray}
Thus
\begin{eqnarray}
&& \sum_n \Lambda_n(0,0,\Psi) = \frac{135}{8} + \frac{945}{8}
- \frac{25,515}{64} \nonumber \\
&& - \frac{945}{16} + \Lambda_6(0,0,\Psi)
- \frac{945}{8} \nonumber \\ &&
- \frac{945}{8} + \frac{25,515}{32} + \frac{25,515}{32}
\nonumber \\ &&
+ \frac{135}{4} - \frac{2835}{8} \nonumber \\ &=&
\frac{93,555}{128} \cos(6 \Psi)
- \frac{2025}{128} \ .
\end{eqnarray}
\end{appendix}


\begin{thebibliography} {99}
\bibitem{TOROID}
The definition of the toroidicity $T$ in Refs. \onlinecite{PRPT,NAS1,NAS2}
is $T= (1/2)\sum_\alpha {\bf r}_\alpha \times {\bf m}_\alpha=(3/2)m_ir \hat k$, 
where ${\bf m}_i$ is the magnetic moment.  One sees that this definition 
yields $T \propto r$, whereas that introduced here has $T \propto r^2$.
Without the factor $r^2$ the toroidal interaction is not simply
proportional to the product of toroidal strengths. More generally,
if the toroidal plaquette has $n>3$ (as in Fig. \ref{FIG1}),
then with the definition proposed here, $T \propto r^{n-1}$ and
$V_{AB} = T_AT_B \Lambda_n(\phi_A, \phi_B, \Psi)/R^{2n+1}.$
\end{thebibliography}
\end{document}